\documentclass{IEEEtran}[12pt]
\IEEEoverridecommandlockouts

\usepackage{cite}
\usepackage{amsmath,amssymb,amsfonts}
\usepackage{algorithmic}
\usepackage{graphicx}
\usepackage{textcomp}
\usepackage{xcolor}
\usepackage{lipsum}
\usepackage{setspace}
\def\BibTeX{{\rm B\kern-.05em{\sc i\kern-.025em b}\kern-.08em
    T\kern-.1667em\lower.7ex\hbox{E}\kern-.125emX}}
\newcommand\numberthis{\addtocounter{equation}{1}\tag{\theequation}}

\title{Detection and Locating Cyber and Physical Stresses in Smart Grids using Graph Signal Processing}

\author{\IEEEauthorblockN{Md Abul Hasnat and Mahshid Rahnamay-Naeini}\\
\IEEEauthorblockA{\textit{Electrical Engineering Department, University of South Florida, Tampa, Florida, USA} \\
hasnat@usf.edu, mahshidr@usf.edu}}
\date{November 2019}

\begin{document}

\maketitle
\begin{abstract}
Smart grids are large and complex cyber physical infrastructures that require real-time monitoring for ensuring the security and reliability of the system.  Monitoring the smart grid involves analyzing continuous data-stream from various measurement devices deployed throughout the system, which are topologically distributed and structurally interrelated. In this paper, graph signal processing (GSP) has been used to represent and analyze the power grid measurement data. It is shown that GSP can enable various analyses for the power grid's structured data and dynamics of its interconnected components. Particularly, the effects of various cyber and physical stresses in the power grid are evaluated and discussed both in the vertex and the graph-frequency domains of the signals. Several techniques for detecting and locating cyber and physical stresses based on GSP techniques have been presented and their performances have been evaluated and compared. The presented study shows that GSP can be a promising approach for analyzing the power grid's data.
\end{abstract}

\begin{IEEEkeywords}
Smart grid security, cyber attack, graph signal processing, local smoothness, vertex-frequency representation.
\end{IEEEkeywords}

\color{black}

\section{Introduction}

Data analysis for the security and reliability of smart grids has attracted lots of attention in the past decade. As the smart grid maintains its proper functioning by continuous acquisition (from an increasing number of sensors deployed in the system) and processing of the measurement data, any attack on the availability and integrity of the data can lead to improper decisions and actions, which may result in instability and failures in the system. For this reason, it is essential to detect and locate anomalies in the smart grid quickly and accurately. 

Different types of cyber attacks can be launched by attackers in smart grids. Among them, the denial of service (DoS) attack \cite{sun18,ding18}, the data-replay attack \cite{sun18,zhang13}, the false data injection attack (FDIA) \cite{esmali15,liang17} have been extensively studied by the researchers. These attacks can, for example, be launched on the supervisory control and data acquisition (SCADA) readings and component status reports as well as on the time-stamped synchrophasor measurements from the phasor measurement units (PMUs). In addition to cyber stresses, physical stresses can also affect the reliability and stability of the system. Examples of such stresses include line and generator failures, and abrupt changes in the loads.

In this paper, Graph Signal Processing (GSP) \cite{ortega18, shuman13} has been exploited for representation and analyses of the smart grid's data for reliability and security purposes. GSP is a fast-growing field, which extends the classical signal processing techniques and tools to irregular graph domain instead of the Euclidean domain. GSP is suitable for analyzing structured data and the dynamics of systems with interconnected components. In this paper, it has been shown that by representing the smart grid data using {\it graph signals}, one can exploit the rich tools that GSP provides to exploit the implicit structures in the smart grid data for security and reliability analyses. Specifically, for the analyses of data from complex networked systems, such as power system, their physical topology as well as the structured interactions (model-based or data-driven interactions) among the components \cite{upama20} are of immense importance. While connectivities and interactions cannot be captured by the classical signal processing approaches, GSP provides a framework to capture such information in graph signals. Since GSP considers the graph structure of the data along with the signals, it is particularly suitable for representing and analyzing data from smart grids.

In this paper, various properties of graph signals both in the vertex domain and the graph-frequency domain have been analyzed. Moreover, several techniques for detecting and locating various cyber and physical stresses in the system have been proposed based on GSP techniques. Specifically, various data integrity attacks (cyber stresses), such as DoS attack, data-replay attack, and false data injection attack as well as failure of a single transmission line (as the physical stress) have been considered in the study. Based on the effects of different stresses on the graph Fourier transform (GFT), the local smoothness, and the vertex-frequency energy distribution of the graph signals, various stress detection and locating techniques are proposed. The performances of the proposed techniques are tested and it is shown that GSP can provide a promising framework for representing power system's data, particularly, for stress detection and locating.

\section{Related Works}

In this section, we briefly review the related work in two categories. In the first category, some of the developments in the area of GSP have been reviewed, and in the second category security studies in smart grids have been discussed. 

Over the last decade, GSP has emerged and extended the concepts of classical signal processing to irregular graph domain. Several works have been published on the interpretation of the frequency domain in the context of graph signals \cite{ortega18, shuman13, sandryhaila14}. The tools and theories built based on these interpretations allow studying graph signals in a new domain with a similar notion to the frequency domain for classical signals. For instance, the relationship between the graph signal frequency and the eigenvalues of the graph Laplacian as well as various concepts related to the graph signal frequency, e.g., global and local smoothness of signals, graph filtering, and modulation of graph signals have been discussed. Moreover, analogous to the joint time-frequency representation of temporal signals, the concept of vertex-frequency analysis of graph signals has been developed and interpreted in \cite{shuman16, stankovic17}. However, unlike the Fourier basis functions, the bases for representing graph signals in the frequency domain, i.e., eigenvectors of the graph Laplacian, are localized in nature. Windowed graph Fourier transform (WGFT) \cite{thanh17} and graph wavelet transform (GWT) \cite{hamm11} have also been introduced. Inspired from the concept of time-frequency energy distributions in the classical signal processing (e.g., Rihaczek energy distribution \cite{stankovic13book}), the work by Stankovi\'{c} et. al. \cite{stankovic18} introduces vertex-frequency energy distributions in the context of graph signal processing. The vertex-frequency energy distributions can be useful for studying the frequency characteristics of the graph signal in a vertex-localized manner. A few works have also been published on the time-vertex signal processing \cite{perra18, perra19}, which treats a time-series associated with each of the vertices of a graph. 

GSP techniques have been used in various applications in the past decade, e.g., sensor networks \cite{wagner05}, biological networks, brain connectivity \cite{golds17}, Electrocardiogram (ECG) signal analysis \cite{sun20}, image, and video processing \cite{egilmez15}. 
Specifically, researchers have shown that GSP can be a prospective field for detecting anomalies in different types of networks \cite{egilmez14, wang14}. However, the application of GSP to smart grids has been limited. For instance, Ramakrishna and Scaglione \cite{ramakrishna19} modeled the voltage phasor measurements in the power grid as the output of the low-pass graph filter in response to the low-rank excitation that comes from the generators. Kroizer et. al. in \cite{kroizer19} approximated the non-linear measurement functions in the power grid as the output of a graph filter and proposed a regularized least-squares estimator for signal recovery based on the inverse of the obtained graph filter.  

Next, we will briefly review security studies in smart grids. Over the past  few decades, cyber security of smart grids has attracted lots of attention \cite{sun18}. Detecting and locating different kinds of cyber attacks  \cite{esmali15, liang17, zhang13} are some of the most important challenges in this context. On the other hand, detecting and locating physical stresses in the grid are also of great importance for the operation and maintenance purpose of the grid. In literature, the detection of cyber attacks and physical stresses have been addressed together under the topic of anomaly detection in the smart grid as well as discussed separately. 

Different techniques for detecting and locating cyber and physical stresses in the smart grid have been proposed in the literature based on both the traditional SCADA measurements as well as the high-frequency PMU measurements. The detection methods based on state estimation of power systems are well suited for the SCADA based static monitoring system while the time-series prediction based methods exploiting the space-time relationship among the states are more applicable to PMU based dynamic system monitoring \cite{cao20}. There are some machine-learning-based as well for the detection of stresses in the smart grid. For the real-time detection of the cyber attacks in smart grids, several works have been emerged. For example, Kurt et. al. \cite{kurt18} proposed a  generalized cumulative sum algorithm for the detection of cyber attack in smart grids in both centralized and distributed manner.  Principle component analysis (PCA) and dimensionality reduction based methods \cite{chen13, mahapatra19} are also being used for the detection of attacks in power systems.  In some works, the spatial and temporal correlations among the states of the power system' components have been exploited to detect and locate cyber attacks and physical stresses in real-time \cite{li19, hasnat19}.  Neural network-based methods for anomaly detection include the works in \cite{basu19, ganjkhani19}. For the detection of line failure in the grid specifically, Hossain and Rahnamay-Naeini \cite{jakir19} proposed a method based on a linear regression method to analyze PMU data. Deng et. al. \cite{deng19} proposed the detection of a single line outage by detecting peaks in frequency signals from the PMUs and locating the outage from the changes in the active power.  

Detection and locating cyber attacks in the smart grid using GSP is new in the power system's literature. Drayer and Routtenberg \cite{drayer18,drayer19} proposed a Graph Fourier Transform (GFT) based detection method for false data injection attacks in smart grids. In their work, it is assumed that the graph signal associated with the bus voltage angles of the power system is smooth and for this reason, the high-frequency components (corresponding to the large eigenvalues of the graph Laplacian) of the graph signal would be insignificant. The existence of the anomalous or false data is proposed to be detected based on the existence of significant high-frequency components. The authors also proposed locating FDIA using graph modulation \cite{drayer191}.  Ramakrishna and Scaglione \cite{ramakrishna191} also utilized their model developed based on GSP in \cite{ramakrishna19} to detect FDIA in the smart grid.

In this work, several techniques for detecting and locating cyber attacks and single transmission line failures in the smart grid have been discussed in the graph-frequency domain as well as in the joint vertex-frequency domain.

\section{GSP Representation of Power system measurements}

\subsection{Preliminaries and Definitions}
The first important definition in GSP is the definition of the graph signal. While in classical signal processing, signals are defined by Euclidean representation of their values; in GSP, the graph signals are defined by the values residing on vertices $\mathcal{V}$ (i.e., $\mathcal{V}=\{v_1, v_2, ..., v_N\}$), which are connected over graph $\mathcal{G}=(\mathcal{V},\mathcal{E})$ with $\mathcal{E}$ representing the set of links (i.e., $\mathcal{E}=\{e_{ij}: (i,j) \in \mathcal{V} \times \mathcal{V}\}$). The graph signal can formally be represented by a vector of values denoted by $\bf{x}$ with size $N$ defined as $ { x}: \mathcal{V} \rightarrow \mathbb{R}$. A graph $\mathcal{G}$ enables capturing the interactions among vertices' variables. Therefore, one of the important steps in defining graph signals is to specify the underlying connectivities among the components, i.e. the graph domain. 

\subsection{Defining graph domain for power grids}

In this paper, our discussion will be limited to the undirected graphs of two types (1) \textit{Bus-vertex} graph: a weighted undirected graph in which buses are considered as the vertices and the transmission lines or the branches are considered as the edges, and (2) \textit{Line-vertex} graph: an unweighted undirected graph in which the transmission lines are considered as the vertices and each edge represents a common bus between a pair of vertices (transmission lines). Since the Bus-vertex graph structure remains unchanged during the cyber attacks, we propose to use this structure for the detection and locating of cyber attacks, whereas the Line-vertex graph structure is more suitable for the study of line failures in the power system, as the Bus-vertex graph changes due to failures. For the rest of the paper, the Bus-vertex graphs will be denoted by $\mathcal{G}$ and the line-vertex graphs will be denoted by $\mathcal{G_L}$.

Note that the above graphs are based on the physical topology of the power system. However, the interactions among the components of the power system can be beyond the physical topology. As such, other methods of constructing a graph domain for power grids can also be used. For instance, the data-driven and electric-distance-based methods discussed in \cite{upama20}, can be used to infer and construct graph domains for power grids beyond their physical connectivities (when needed depending on the analyses of interest).

In this paper, the geographical distance between buses $i$ and $j$ is denoted by $d_{ij}$ and the weight corresponding to the edge $e_{ij}$ in the bus-vertex graph $\mathcal{G}$ is defined as $w_{ij} = \frac{1}{d_{ij}}$, if there is an edge between node $i$ and node $j$ (i.e., $e_{ij}=1$) and $w_{ij}=0$, otherwise (if there is no edge between node $i$ and node $j$, i.e., $e_{ij}=0$). For the line-vertex graph $\mathcal{G_L}$, the weights of the edges are considered as: $w_{ij} = 1$, if $e_{ij} = 1$, and $w_{ij}=0$, otherwise. 

Another important definition related to the graph is its Laplacian matrix $\textbf{L}$, where its element $l_{ij}$ is defined as:

 \[ l_{ij} = \begin{cases} 
          \sum_{j=1}^N w_{ij}, & if, \ \ i=j \\
          -w_{ij} & if, \ \ i \neq j \\
       \end{cases}
    \]
Since, the graph Laplacian, $\textbf{L}$ is a real and symmetric matrix, it has real and non-negative eigenvalues corresponding to the orthonormal set of eigenvectors.  The Laplacian matrix of the graph will be used later in defining the frequency domain representation of graph signals.

\subsection{Representation of Power System Measurements as Graph Signals: Vertex Domain Representation}

The graph signal associated with the graph, $\mathcal{G}$, can be defined as a mapping of the vertices into a vector of real numbers with size $N$, $x:\mathcal{V} \rightarrow \mathbb{R}$. Here, $x(v_n)$ is the value of the graph signal at vertex $v_n$. However, for simplicity, this signal will be denoted by $x(n)$ instead of $x(v_n)$ for the rest of the paper. 

\begin{figure}[h]
  \centering
  \vspace{-0.2in}
  \hspace{-0.3in}
  \begin{tabular}{c}
    \includegraphics[height=2.2in]{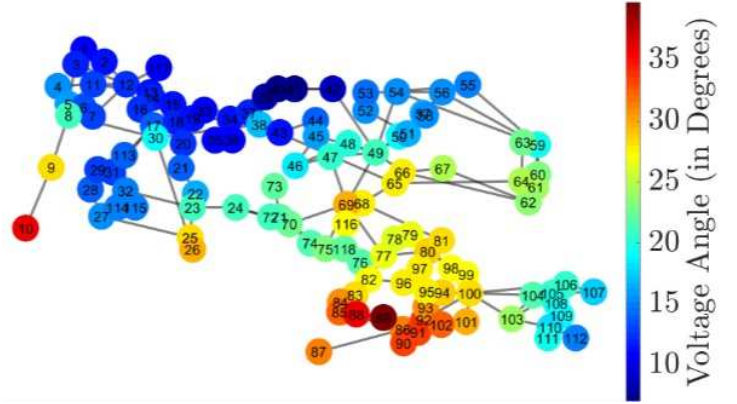}\\
  \end{tabular}
  \vspace{-0.2cm}
  \caption{Voltage angle measurements at a particular time instance as a Graph Signal on the IEEE 118 bus system.}
\label{gsp118}
\end{figure}

In this paper, we consider the measurement values associated with each vertex (bus voltage angles for $\mathcal{G}$ and real-power flow or line current angle for $\mathcal{G_L}$) at a time instance as a graph signal. These measurement values can be obtained from the SCADA readings, or from the PMU measurement at a specific time instance (assuming PMUs are mounted on every bus in the system). Fig. \ref{gsp118} illustrates the graph signal of the voltage angles of all the buses for the IEEE 118 bus system \cite{118bus}. On the other hand, $x_L(n)$ is the graph signal associated with $\mathcal{G_L}$, which is corresponding to the current measurements in the transmission lines of the system.

The measurement values at different time instances can be modeled as time-series associated with each vertex. The resultant graph signal becomes also a function of time, i.e., a time-varying graph signal that has been discussed in detail in subsequent subsections.  

\subsection{Spectral Characteristics of Power Grid's Graph Signal - Graph-Frequency Domain}

\begin{figure}[h]
  \centering
  \hspace{0in}
  \begin{tabular}{c}
    \includegraphics[height=2.5in,width=1\columnwidth, trim =5cm 0cm 0cm 0cm, clip]{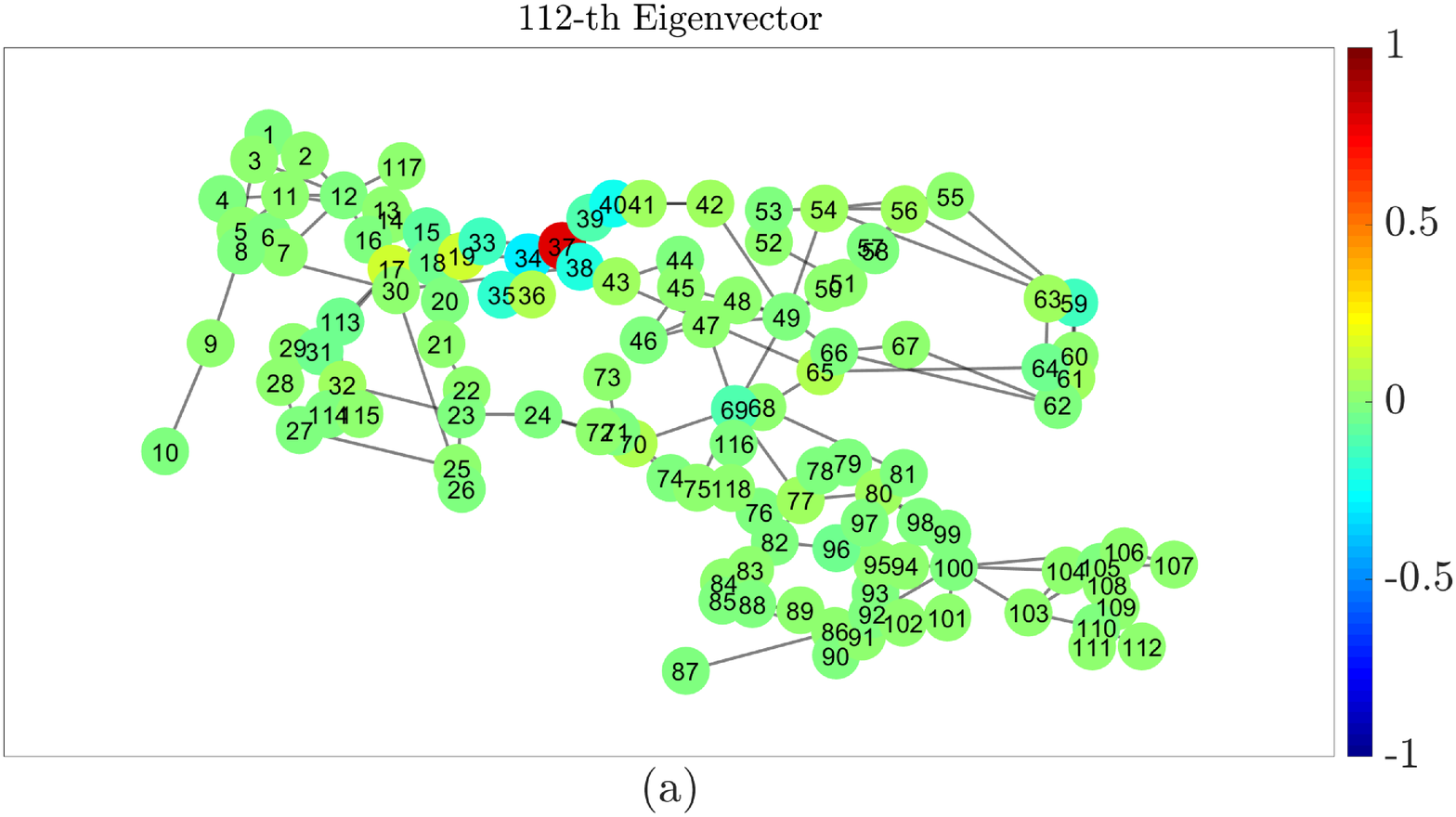}\\

\includegraphics[height=2.5in,width=1\columnwidth, trim =5cm 0cm 0cm 0cm, clip]{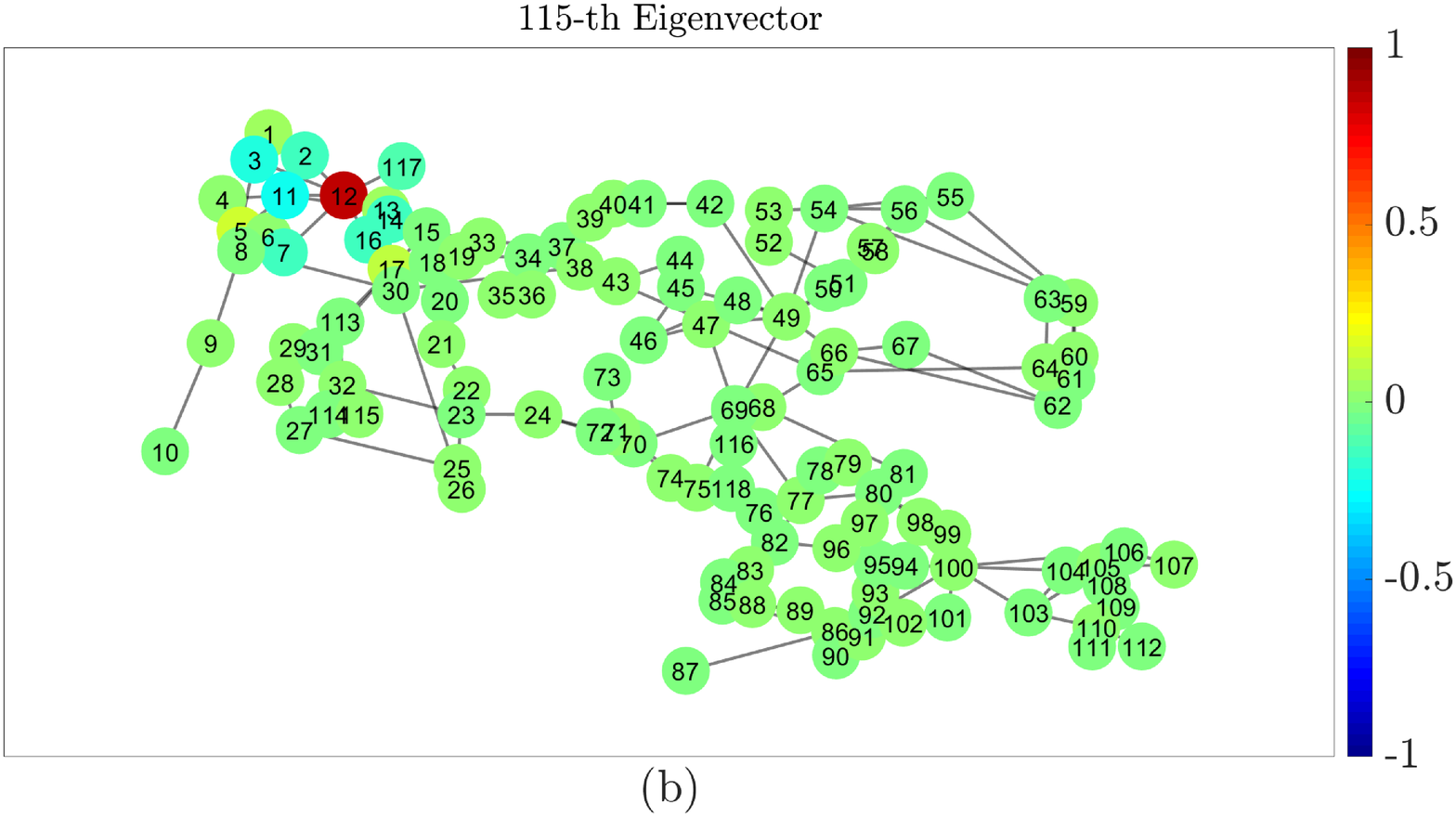}
  \vspace{-0.2cm}  
  \end{tabular}
  \caption{Two of the eigenvectors for IEEE 118 bus systems Graph. The two eigenvectors are localized around two different vertices.}
  \vspace{-0.4cm}
\label{eigvec118}
\end{figure}

In classical signal processing, the concept of frequency for temporal signals is explained with Fourier transform, which is a linear transform with basis function $e^{j \omega t}$, where $\omega=2\pi f$ and $f$ is the frequency variable. In fact, $e^{j \omega t}$ is the eigenfunction of the one dimensional Laplacian operator $\Delta$, i.e., $\Delta(e^{j\omega t}) = - \omega^2 e^{j\omega t}$. The Fourier transform of a temporal signal $x(t)$ is:

\begin{equation}
    \hat{X}(\omega) = \int_{t=-\infty}^{\infty} {x(t)(e^{j\omega t})^* dt}, \  \  (Analysis \ \ equation)
\end{equation}

\begin{equation}
    x(t) = \int_{\omega=-\infty}^{\infty} {\frac{1}{2\pi}\hat{X}(\omega)(e^{j\omega t})} d\omega,\ \  (Synthesis \ \ equation).
\end{equation}

Analogous to the concept of Fourier transform and frequency domain representation of the signal, the basis functions for the graph Fourier transform are considered as the eigenvectors denoted by $u_k(n)$ of the graph Laplacian $\textbf{L}$ (defined in Section III.B), where subscript $k$ denotes the $k-$th eigenvector and $n$ is the index of $n-$th node in the graph $\mathcal{G}$.   The corresponding eigenvalues to these eigenvectors are denoted by ($\lambda_{k}$) and are considered as the graph-frequencies.

The graph Fourier transform (GFT) of a graph signal $x(n)$ is:
\begin{equation}
     \hat{X}(\lambda_{k}) = \sum_{n=1}^{N} {x(n)u_k(n)}, \  \  (Analysis \ \ equation) \\
\end{equation}
and the inverse graph Fourier transform (IGFT) is:
\begin{equation}
     {x}(n) = \sum_{k=1}^{N} {\hat{X}(\lambda_{k})u_k(n)}, \  \  (Synthesis \ \ equation) \\
\end{equation}
where $\lambda_k$ and $u_k$ are, respectively, the $k-$th eigenvalue and the $k-$th eigenvector of $\textbf{L}$, where $ 0 = \lambda_1 < \lambda_2 < \lambda_3 < ... < \lambda_N$.  The first eigenvalue $\lambda_1=0$ is analogous to the zero-frequency (DC component) in the case of temporal signals. The eigenvectors with lower eigenvalues correspond to the lower frequency components with less variation of the values of the signal as local neighborhoods. The higher eigenvectors (larger $k$) correspond to the high-frequency components that have a higher rate of changes in the node-to-node values. In contrast to the classical Fourier transforms basic functions, the graph Laplacian eigenvectors are localized in the vertex domain. For example, Fig. \ref{eigvec118} illustrates two eigenvectors of the graph structure corresponding to the IEEE 118 bus system that are localized around two different locations in the graph.

\subsection{Global and Local Smoothness of Graph Signals}

The smoothness measure of a signal quantifies how rapidly the values of the signal changes. In a graph signal, the smoothness characterizes the variation of the signal over graph neighborhoods, i.e., from each vertex to its neighboring vertices. The global smoothness signifies the aggregated variations in the signal while local smoothness signifies variation in the vicinity of each vertex.

\paragraph{Global Smoothness} 
The global smoothness of a graph signal $x(n)$ is defined as:
\begin{equation}
s_{Global} = \frac{{\textbf{x}}^T\textbf{L}\textbf{x}}{{\textbf{x}}^T\textbf{x}},
\end{equation}
where $\textbf{x}$ is the vector representation of the graph signal, $x(n)$.  The faster the graph signal changes  from vertex to vertex, the larger the value of  $s_{Global} $.

\paragraph{Local Smoothness}

The local smoothness of the graph signal $x(n)$ at vertex $n$ is defined as:
\begin{equation}\label{local}
s(n) = \frac{l_{\textbf{x}}(n)}{x(n)},  \ \ x(n)\neq 0,
\end{equation}
where $l_{\textbf{x}}(n)$ is the $n-$th element of the vector, $\textbf{L}\textbf{x}$. In equation (\ref{local}), $s(n)$ signifies how fast the values of the graph signal $x(n)$ changes from vertex to vertex in the vicinity of the $n-$th vertex. The work by Dakovi\'c et al. presented in \cite{dakovic} shows that the concept of local smoothness in the graph signal is analogous to the concept of instantaneous frequency in classical signal processing.

\subsection{The Joint Vertex-Frequency Representations}

In classical signal processing, the joint time-frequency representation of signals (e.g., spectrum, windowed Fourier transform, wavelets, etc.) are used for the time-localization of a particular frequency component. The joint vertex-frequency representations serve the similar purpose for graph signals. There are different approaches for the localization of the frequency components in the literature. For example, Stankovi\'c et. al. \cite{stankovic17} propose localized vertex spectrum (LVS) of graph signal $x(n)$ as:
\begin{equation}
    LVS_x(n,\lambda_k) = \sum_{m=1}^N {x(m)h(n-m)u_k(m)}, 
\end{equation}
where $h(n)$ is the window function. This approach has a major drawback of being dependent on the width and the characteristics of the window function. Instead, for improving the localization of the signal energy in the joint vertex-frequency domain, the vertex-frequency energy distribution is introduced in \cite{stankovic18}, which does not require any window. The Vertex-frequency energy distribution $E(n,k)$ is calculated from the graph signal using the equation:
\begin{equation}\label{vfed}
         E(n,k) = \sum_{m=1}^N x(n)x(m)u_k(m)u_k(n). 
\end{equation}

However, the smoothed version of the vertex-frequency energy distribution makes the vertex-frequency representation less sensitive to changes and disturbances.

\subsection{Time-Varying Graph Signals}

In our previous discussions, we have only considered the graph signal at a single time instant. However, in dynamic systems, such as power grids, values of the signal at each node vary in time. For instance, the bus voltage measurements in power grids change in time because of changes in load demand and other changes in the power system. As a result, the graph signal $x(n)$ changes in time. Therefore, a time-varying graph signal can be thought of as a function of both vertex and time and can be denoted by $x(n,t)$. It is worth mentioning that, the graph itself, $\mathcal{G}$, i.e., the vertices, edges, weights are not changing with time in this case. If the topology of a graph changes with time, it results in a dynamic graph, which is out of the scope of this paper.

For time-varying graph signal $x(n,t)$, the spectral representations as well as the global and local smoothness of the graph signals also change with time. In this paper, the $k-$th eigenvalue, the $k-$th eigenvector, the GFT, and the vertex-frequency energy distribution at time $t$ will be denoted by $\lambda_k(t), u_k(t), \hat{X}(\lambda_k,t),$ and $E(n,k,t)$, respectively. The local smoothness of vertex $n$ at time $t$ will be denoted as $s(n,t)$.

\section{GSP for Detecting and Locating cyber attacks in smart grids}    

\subsection{Modeling cyber attacks in graph signals}

For modeling cyber attacks in graph signal domain, let us consider that a set of vertices, $\mathcal{V_A} \subset \mathcal{V}$ is under attack within the time interval $t_{start}$ to $t_{end}$. The corrupted signal can be expressed as follows:
\begin{equation}
    x(n_A,t) = c(t), \ \ for \ \ t_{start} \leq t \leq t_{end},  \label{model}  
\end{equation}
where $n_A \in \mathcal{V_A}$. In equation (\ref{model}), $c(t)$ represents the corrupted signal under cyber attack in general case. Special types of cyber attacks can be modeled by different considerations over $c(t)$. For instance, in case of DoS attack, one can consider $c(t) = 0$. For data replay attack, $c(t) \in \{x(n,t_p)\}$, where $t_p<t_{start}$.  For other false data injection attacks (FDIA), $c(t)$ is designed intelligently using different techniques usually to hide the attack. 

\vspace{-0.2cm}

\subsection{Detection of cyber attacks using GFT}
\begin{figure}[h]
  \centering
  \vspace{-0.1in}
  \begin{tabular}{c}
    \includegraphics[height=2in,width=1\columnwidth, trim =2cm 0cm 2cm 1cm, clip]{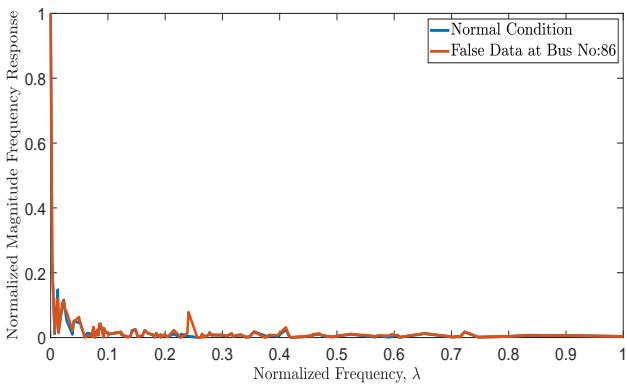}
  \end{tabular}
  \caption{GFT Magnitude Response for IEEE 118 bus system: Emphasized high graph-frequency components can be observed in case of false data injection.}
\label{gft}
\end{figure}

In this section, it is shown that stresses in the power grid can be detected from the GFT coefficients of the signal. In general, the low-frequency components are prominent for the bus voltage angle graph signals because of the smooth changes of bus-to-bus values due to the power flow dynamics. The GFT coefficient magnitudes with respect to the normalized graph-frequencies (i.e., $\hat{\lambda}_k=\frac{\lambda_k-min_i\{\lambda_i\}}{max_i\{\lambda_i\}-min_i\{\lambda_i\}}$) are illustrated in Fig. \ref{gft} for a bus-voltage angle graph signal defined on the graph of the IEEE 118 bus system under normal conditions as well as under a false data injection attack. It can be observed that the magnitude of the high-frequency components is larger in the second case. This property can be exploited for detection of anomalies in the measurement data. A parameter $\gamma(t)$ is introduced to quantify the amount of high graph-frequency components corresponding to a graph signal $x(n,t)$ at the time instant $t$ as follows:
\begin{equation}
    \gamma(t) = \sum_k |\hat{X}(\hat{\lambda}_k,t)H(\hat{\lambda}_k)| , 
\end{equation}
where $H(\lambda)$ is a high-pass graph filter expressed by the following frequency response:
\begin{equation}
    H(\lambda) =
  \begin{cases}
                                   0 & \text{if $\lambda \leq {\lambda}_c$} \\
                                   1 & \text{if $\lambda > {\lambda}_c$}, \\
  \end{cases}
\end{equation}where ${\lambda_c}$ is the cut-off graph-frequency.
For detecting cyber and physical stresses, we estimate the probability distribution of the parameter $\gamma$ in normal conditions from the historical data (assuming $\gamma$ is a stationary random variable and denote its distribution as $p_{\gamma}(\zeta)$). For a certain time instant $t$, a stress is declared if the likelihood of $\gamma(t)$ corresponding to the distribution is less than a certain threshold $\theta_{\gamma}$, (i.e., $p_{\gamma}(\gamma(t))<\theta_{\gamma}$). The threshold $\theta_{\gamma}$ is selected empirically considering the tail probabilities of  $p_{\gamma}(\zeta)$.  Although this method detects cyber attacks reasonably well, the major drawback of this method is that it cannot provide any information about the location of the attack.  

\vspace{-0.1cm}

\subsection{Detecting and locating stresses using smoothness of graph signals}
\begin{figure}[h]
  \centering
    \includegraphics[height=2.5in,width=\columnwidth, trim =3cm 0cm 0cm 0cm, clip]{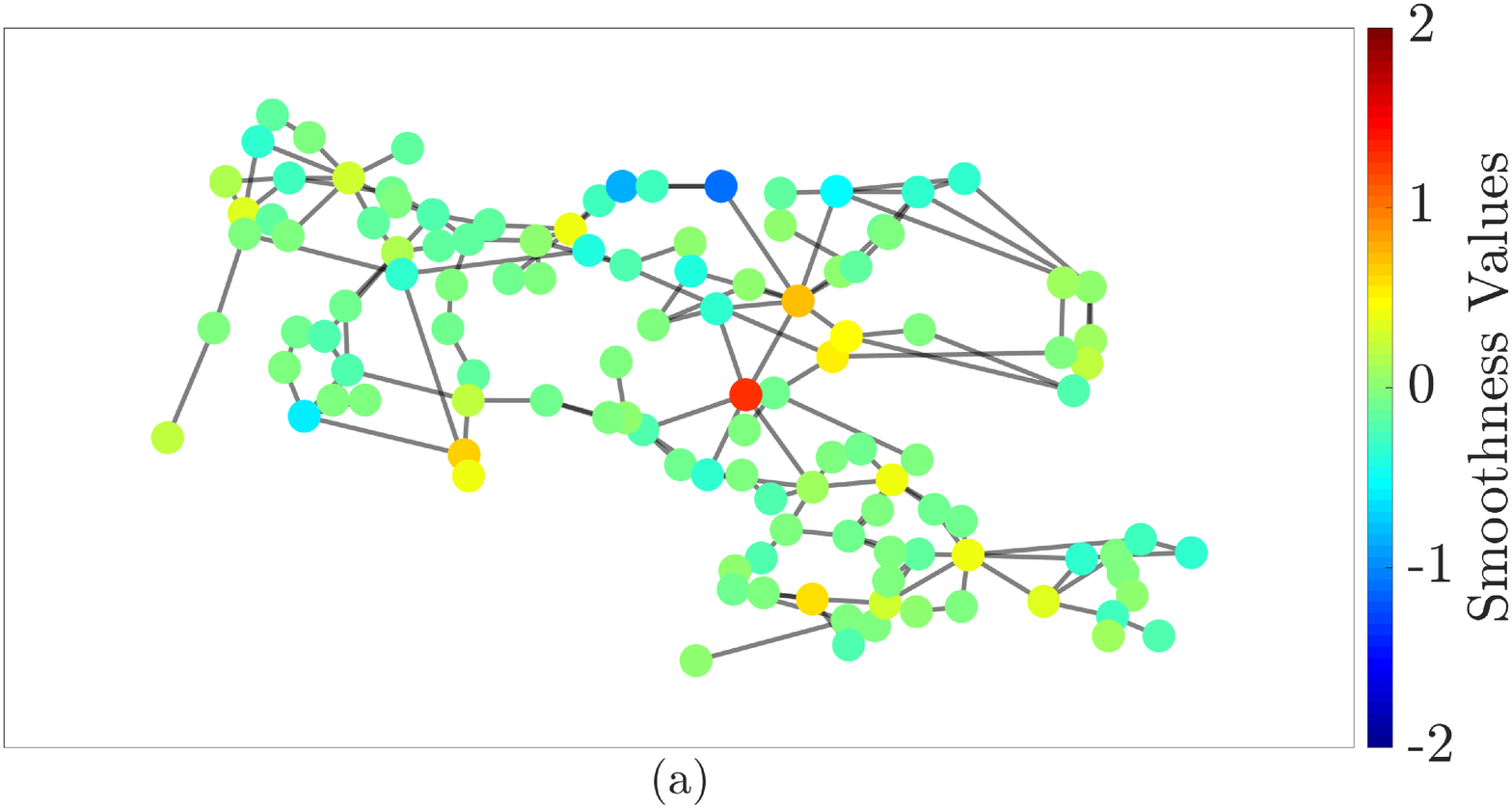}\\
    \vspace{-0.3cm}
    \includegraphics[height=2.5in,width=\columnwidth, trim =3cm 0cm 0cm 0cm, clip]{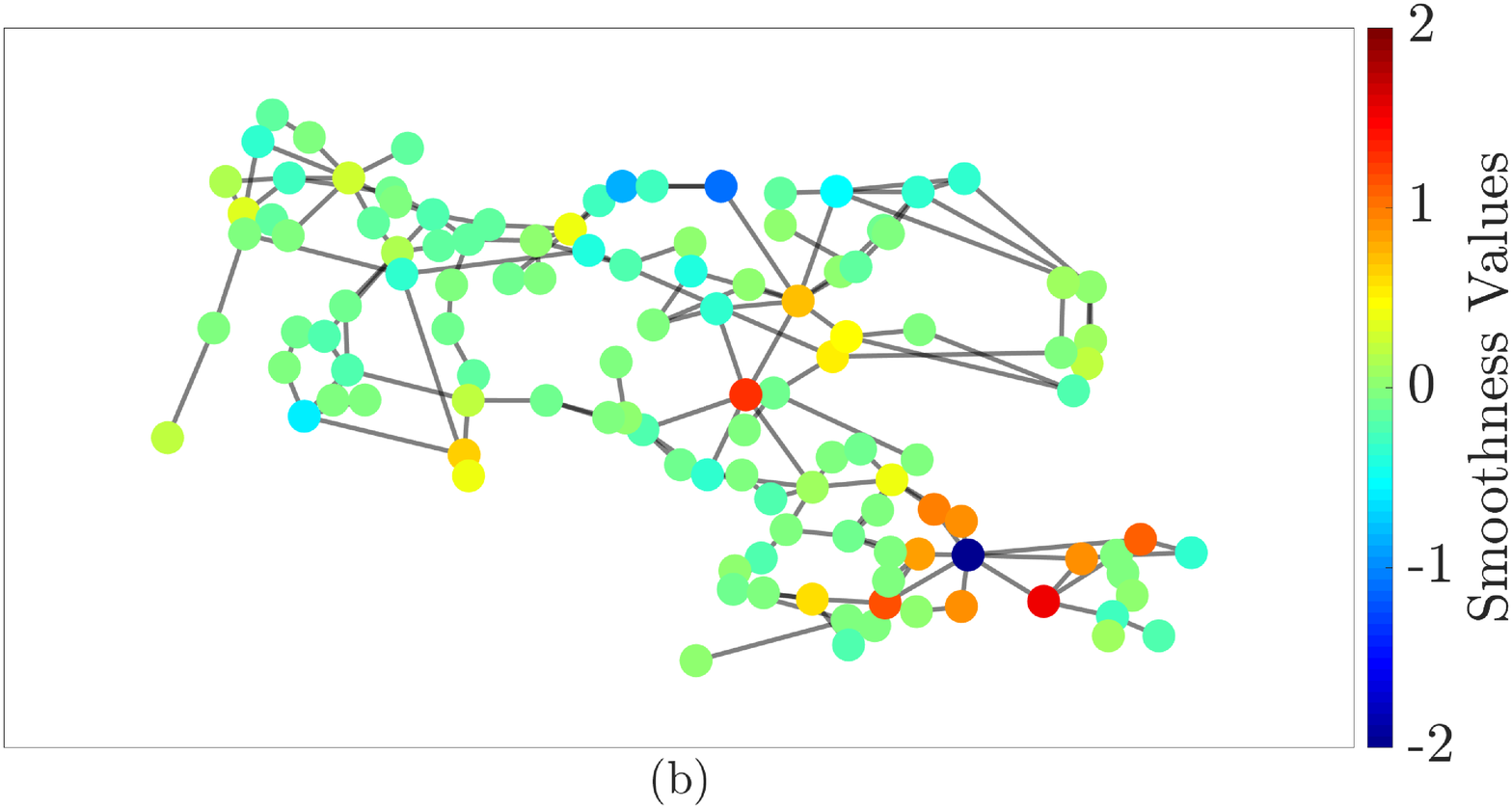}
    \vspace{-1cm}
    \caption{Local Smoothness of the vertices of the IEEE 118 bus system: (a) Normal Condition, (b) DoS Attack at bus number 100.}
    \vspace{-0.4cm}
\label{locsmooth}
\end{figure}

Fig. \ref{locsmooth} illustrates the local smoothness of the vertices of the IEEE 118 bus system corresponding to the bus-vertex graph $\mathcal{G}$ and graph signal $x(n)$ in the normal condition (Fig. \ref{locsmooth}(a)) as well as under DoS attack at bus number $100$. It can be observed that the local smoothness of the vertices in the vicinity of vertex number $100$ have changed significantly. Here, we propose to exploit this effect on the local smoothness of the vertices to detect and locate cyber attacks. For detecting and locating cyber attacks and line tripping, the historical data have been used to estimate the probability distribution of the local smoothness of the $n-$th vertex  $p_{s_n}(\zeta)$ under normal conditions. At time instant $t$,  if the likelihood of $s(n,t)$ is less than a certain threshold $\theta_{s_n}$ (i.e., $p_{s_n}(s(n,t))<\theta_{s_n}$), a stress is declared at vertex $n$.

A similar technique can be applied on the graph signal $x_L(n_L)$ representing real-power flow through transmission line $n_L$ associated with the line-vertex graph $\mathcal{G_L}$  to detect and locate single line failure. 

\subsection{Detecting and locating stresses using vertex-frequency energy distribution} \label{method_cyber}

In this work, it is shown that the vertex-frequency energy distribution of the time-varying graph signal $x(n,t)$ associated with the bus-vertex graph $\mathcal{G}$ can be used to detect and locate anomalous data in smart grids. Containing the topological and the spectral information simultaneously,  the vertex-frequency energy distribution makes itself more suitable for detecting and locating anomalies in complex networks. Moreover, due to the better concentration of signal energy compared to the linear joint vertex-frequency representations, it serves better for locating cyber attack. 

Let us consider the time-vertex graph signal $x(n,t)$, which can also be thought as a set of time series each associated with the vertices. Let $t_A$ be the time instant at which the cyber attack is introduced into the grid. $x(n,t_A-\epsilon)$ is the vertex-domain graph signal just before the attack (under normal conditions), whereas $x(n,t_A+\epsilon)$ is the vertex-domain graph signal just after the attack (under a cyber attack) and $\epsilon$ is a small real value. $E(n,k,t_A-\epsilon)$ and $E(n,k,t_A+\epsilon)$ are the vertex-frequency energy distributions corresponding to the graph signals $x(n,t_A-\epsilon)$ and $x(n,t_A+\epsilon)$, respectively.

\begin{figure}[h]
  \centering
  \hspace{-0.5in}
  \begin{tabular}{c}
 \includegraphics[height=2.4in,width=\columnwidth, trim =4cm 0cm 8cm 0cm, clip]{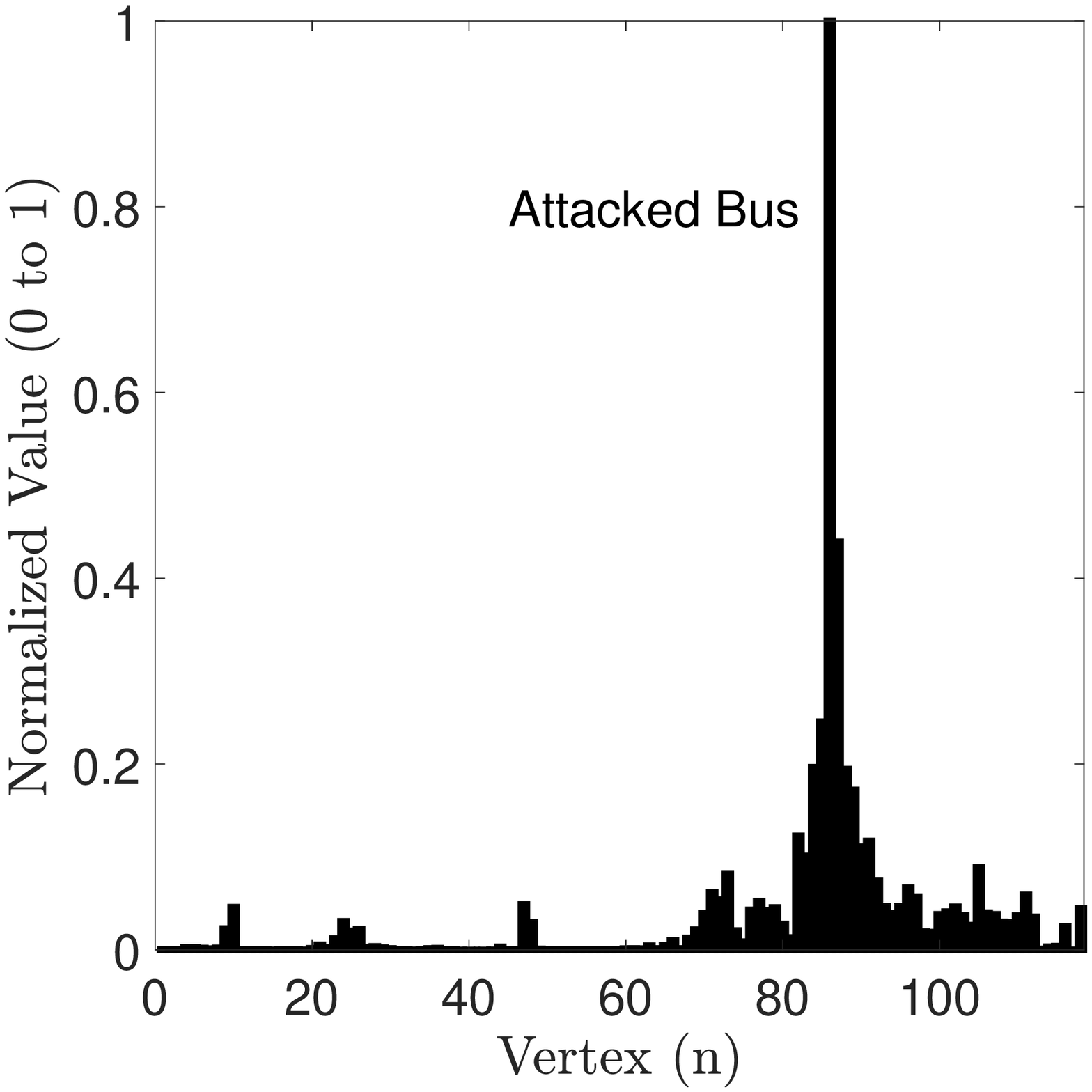}
    
\end{tabular}
    \caption{ Marginalization over frequency components for locating vertex (bus), $\sum_{k=1}^N |E(n,k,t_a+\epsilon)-E(n,k,t_a-\epsilon)|$. For $n=86$, the largest value is obtained which indicates a stress at vertex (bus) $86$.}
    \vspace{-0.5cm}
\label{enk118}
\end{figure}

Cyber attacks (either DoS attacks or false data injection attacks) involve abrupt and abnormal changes in the time-vertex graph signal $x(n,t)$, which also affect the graph-spectral characteristics of the graph signal at that time instant i.e., $E(n,k,t_A+\epsilon)$. Hence, the vertex-frequency energy distributions before and after the cyber attack would have certain differences. Our proposition is to utilize the difference between these two, i.e., $|E(n,k,t_A+\epsilon)-E(n,k,t_A-\epsilon)|$, to detect and locate cyber attacks. By marginalizing the difference distribution over the frequency axis, a comparatively larger value has been obtained for the compromised vertices,  e.g., vertex 86 of IEEE 118 bus system in Fig. \ref{enk118}.

Although the joint vertex-frequency energy distribution provides important insights and justifications for detecting and locating cyber attacks in the power grid, the direct implementation of this technique on $|E(n,k,t_A+\epsilon)-E(n,k,t_A-\epsilon)|$, faces the challenge of selecting the proper threshold of detection in the joint-vertex frequency domain. For this reason, the instantaneous vertex-frequency energy distribution $E(n,k,t)$ is directly marginalized over the spectral index $k$ to obtain the energy distributions with respect to the vertex indices at a certain time:
\begin{equation}\label{margin}
    \sum_{k=1}^{N} E(n,k,t) = |x(n,t)|^2,
\end{equation}      
and set the threshold of detection over the distribution of $|x(n,t)|^2$. We consider that for the voltage angle measurements the values of $x(n,t)$ are normally distributed for each $n$ and $t$ with a mean value $\mu_{n,t}$ and standard deviation $\sigma_n$. Therefore, the probability distribution of $|x(n,t)|^2$ is a piece-wise gamma distribution with the parameters, $\mu_{n,t}$ and $\sigma_n$, for each $n$ and $t$: 
 \begin{align*}
 p_{n,t}(y) &= \frac{1}{2\sigma_n\sqrt{2\pi y}} [e^{-(\frac{y-\mu_{n,t}}{2{\sigma_n}^2})} u(y-\mu_{n,t}) +\\ & e^{-(\frac{\mu_{n,t}-y}{2{\sigma_n}^2})} u(\mu_{n,t}-y)],\numberthis \label{pdf}
\end{align*}
where $u(y)$ is the unit step function.

Here, $\mu_{n,t}$  is approximated by the sample mean within a small interval $[t-\delta, t]$. Within that small interval $x(n,t)$ has been considered to be wide sense stationary:
\begin{equation}
    \mu_{n,t} = \int_{t-\delta}^t {x(n,\tau)} d{\tau}. 
\end{equation}
For all the vertices $n$, the standard deviation ${\sigma}_n$ is obtained from the past historical data. For each time instant and each bus, the likelihood of $|x(n,t)|^2$ is calculated using Equation (\ref{pdf}). If the likelihood is less than a certain threshold $\theta_{x^2,n}$ for any $|x(n,t)|^2$, an attack is declared and bus number $n$ is considered as the compromised bus. For the detection and locating of the physical failure, a similar method can be applied on the line-vertex graph $\mathcal{G_L}$ using the graph signal $x_L(n_L,t)$, which represents the angle of the line current through branch $n_L$.
\vspace{-0.1cm}

\color{black}

\section{Simulations and Results}

The IEEE 118 bus system has been used for the evaluation of the proposed techniques. The power flow solutions have obtained using MATPOWER 6.0 \cite{MATPOWER}. For time series associated with the graph vertices, the time-varying load patterns from the New York Independent System Operator (NYISO) \cite{nyiso} have been used.   

\vspace{-0.3cm}
\begin{figure}[h]
  \centering
  \hspace{0in}
  \begin{tabular}{c}
    \includegraphics[height=1.8in, width=3.25in, trim =0cm 0cm 2cm 0cm, clip]{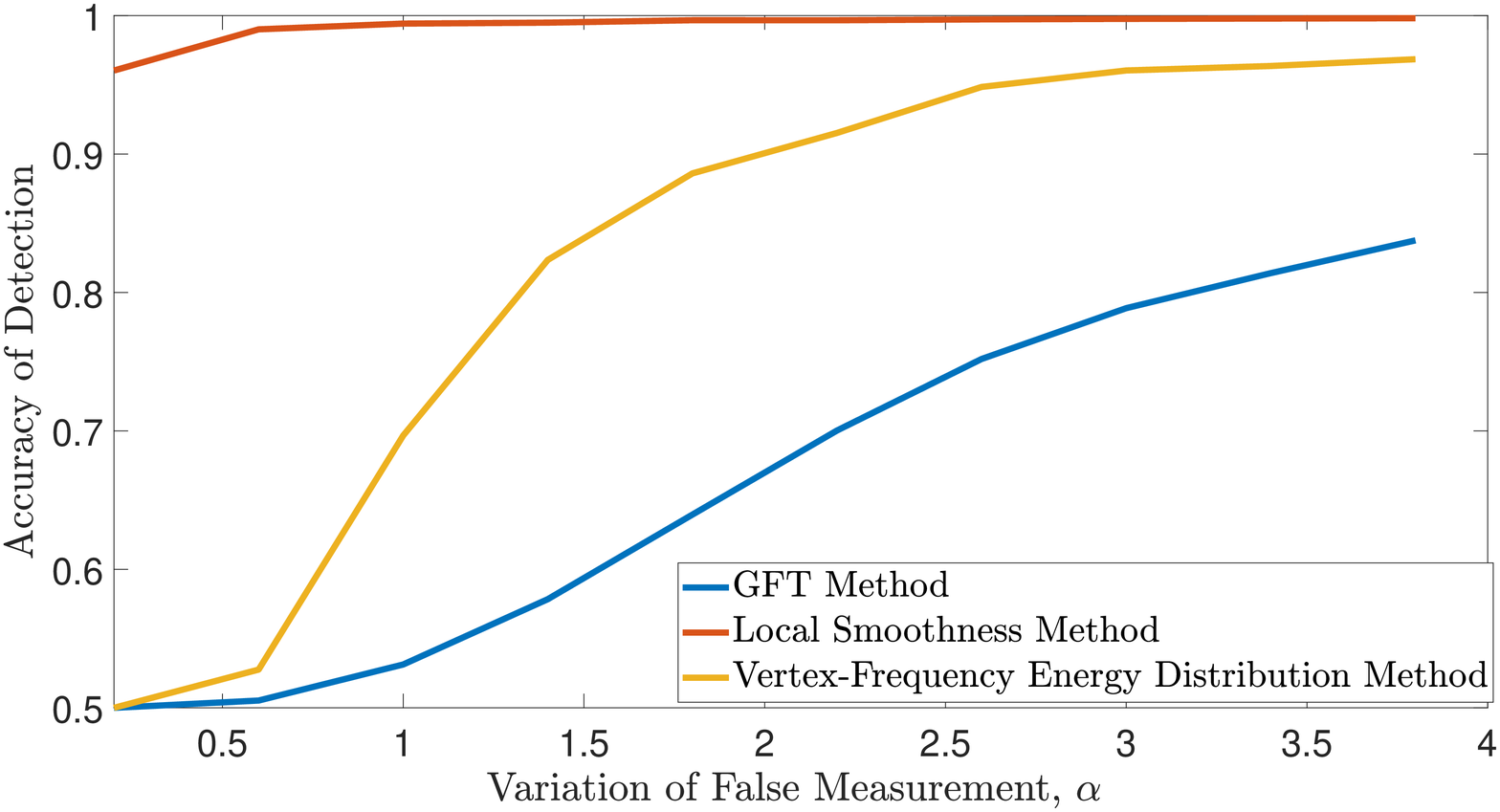}
  \end{tabular}
  \caption{Detection accuracy with respect to $\alpha.$}
\label{performance}
\end{figure}
For the evaluation of the performance of the detection and locating techniques for cyber attacks, the anomalous data associated with the $n_A-$th bus, ($n_A \in \mathcal{V_A}$, where $\mathcal{V_A} \subset \mathcal{V}$ is the set of vertices with falsified measurements) have been designed as follows:
\begin{equation}
    {\theta_{n_A}}_{False} = {\theta_{n_A}} + (-1)^d \alpha.u.Range(\theta_{n_A}),
\end{equation}
where ${\theta_{n_A}}$ is the true measurement associated with the $n_A-$th bus, ${\theta_{n_A}}_{False}$ is the falsified measurement, $d \in \{0,1\}$ is a binary random variable, $u \in [0,1]$ is a random number,   $Range(\theta_{n_A})$ is the range of the true measurement at the $n-$th bus obtained from the simulation data, and $\alpha$ is the random parameter representing the amount of change in the measurement values. As $\alpha$ increases, the detection of attack becomes easier. In this paper, the performance of cyber attack detection using the three proposed methods have been analyzed for various values of $\alpha$.  Specifically, $3,000$ random scenarios have been generated for each value of $\alpha$ among which some of the scenarios are normal and in the other scenarios, there are one or more cyber attacks. The number of attacks, the time instant of the attacks, and the injected false value during the attack are also selected randomly. The accuracy of cyber attack detection for different $\alpha$ values in the three methods have been illustrated in Fig. \ref{performance}. Both the value of $\theta_{\gamma}$ and $\theta_{s_n}$ are selected as $0.005$. It can be observed that the local smoothness method performs better than the other two methods. Although the local smoothness method is the best among the three for detecting cyber attacks, the vertex-frequency energy distribution method outperforms this method in terms of the locating rate. For example, when $\alpha=4$, the locating rate in the local smoothness method is $0.85$ whereas it is $0.91$ in the vertex-frequency energy distribution method. 

For the performance evaluation of the detection and locating techniques for a single line failure, $3,000$ random scenarios have been created among which some scenarios are normal and some scenarios involve single line failures. The tripped branches are selected randomly. The accuracy of detection for single line tripping in the vertex-frequency energy distribution method is $0.80$ with a false positive rate of $0.25$ whereas in the local smoothness method the detection rate is $0.93$ with a false positive rate of only $0.03$. The locating rate within 2-hop distance and 3-hop distance in the local smoothness method are, respectively, $0.55$ and $0.65$.

\section{Conclusion}
In this work, graph signal processing is utilized to represent and analyze the power grid's measurement data for reliability and security evaluation of the system under various stresses. The physical structure of the power grid has been used to define the graph domain with the measurements associated with the grid as the graph signals. The effects of the cyber and physical stresses on the graph signals have been studied in the vertex domain, graph-frequency domain, and joint vertex-frequency domain of the signals. Based on the observations from the effects of stresses, three techniques for detecting and locating cyber and physical stresses based on the graph Fourier transform,  the local smoothness of graph signals, and the vertex-frequency energy distributions are proposed. The performance of the proposed methods has been evaluated using various scenarios of cyber and physical stresses. It has been shown that the graph signal processing techniques can provide good performance for detecting and locating stresses in the system.

\end{document}